\documentclass[aip,cha,amsmath,amssymb,reprint]{revtex4-1}
\usepackage{color}
\usepackage{graphicx}
\usepackage{overpic}
\usepackage{dcolumn}
\usepackage{bm}
\usepackage[mathlines]{lineno}
\usepackage{amsfonts}
\usepackage{amsmath}
\usepackage{sidecap}
\usepackage{float}
\usepackage{subfigure}
\usepackage{parskip}
\usepackage{grffile}

\usepackage{color}
\usepackage{hyperref}
\usepackage{hhline}
\usepackage{mathtools}
\usepackage{comment}
\usepackage[normalem]{ulem}
\makeatletter
\def\@eqnnum{{\normalsize \normalcolor (\theequation)}}
 \makeatother
\begin{document}
\title{Jayesh M\"{o}bius} 

\title{Low dimensional Watanabe-Strogatz approach for Kuramoto oscillators with higher-order interactions}

\author{Jayesh C. Jain and Sarika Jalan}
\email{Corresponding Author: sarika@iiti.ac.in} \affiliation{Complex Systems Lab, Department of Physics, Indian Institute of Technology Indore, Khandwa Road, Simrol, Indore-453552, India}

 \begin{abstract}
      Watanabe-Strogatz theory provides a low-dimensional description of identical Kuramoto oscillators via the framework of the M\"{o}bius transformation. Here, using the Watanabe-Strogatz theory, we provide a unifying description for a broad class of identical Kuramoto oscillator models with pairwise and higher-order interactions and their corresponding higher harmonics. We show that the dynamics of the Watanabe-Strogatz parameters are the same as those of the mean-field parameters. Additionally, the poles of the M\"{o}bius transformation serve as basin boundaries for both global and cluster synchronization in the models discussed here. We present numerical simulations that illustrate how the basins boundaries evolve for these extended models.
        
    

\end{abstract}

\maketitle

\begin{quotation}
    Synchronization is a universal phenomenon observed in widely distributed real-world systems, from the chirping of birds to neuronal activity. One of the frameworks for understanding the origin and prediction of synchronization phenomena in interacting dynamical units is the Kuramoto model. The model has been extensively studied using the Ott-Antonsen (OA) Approach, a dimensional reduction scheme with a special ansatz. In this article, we analyze the model using a more general and exact approach called the Watanabe-Strogatz (WS) approach. Unlike the OA approach, the WS approach can be used to describe a system of finite oscillators and therefore can be applied to model real-world systems which are finite in size. Moreover, the WS approach can be used for a broad class of Kuramoto models that follow the Riccati equation, allowing to monitor evolutions of basin boundaries for these models.
\end{quotation}

\section{Introduction}
The Riccati equation named after the mathematician J. F. Riccati\cite{riccati} has a general form
 \begin{equation}
     \frac{dy}{dx} = a(x) y^2 + b(x) y + c(x),
     \nonumber
 \end{equation}
 where $a,b$ and $c$ are real value functions, while $x$ is an independent variable. 
 The equation naturally arises in many fields of physics and mathematics. The complex form of the Riccati equation has a wide variety of uses in quantum mechanics, cosmology, and thermodynamics. Few examples include the time-dependent Schr\"{o}dinger equation \cite{Schuch_2014},\cite{rosu}, Bose-Einstein condensate \cite{BEC}, the Friedmann equations with self-interacting matter field in cosmology \cite{cosmology}, and nonlinear systems including the Kuramoto model. 
 Other known applications of the Riccati equation pertain to stochastic realization theory, financial mathematics, and network synthesis \cite{ndiaye}.
Further, the Kuramoto model stands as a simple mathematical framework to study synchronization phenomena in a range of complex real-world systems, such as Josephson Junctions, lasers, functional connectivity of the human brain, circadian rhythms, neuronal oscillations, power networks, smart grids, etc. \cite{kuramoto, Review, Priyanka2,JI, tanaka}. 
Ott and Antonsen provided a method for reducing the dimensionality of $N$ coupled phase oscillators at the limit $N\rightarrow \infty$ \cite{OA}. This approach assumes a particular ansatz known as the OA ansatz, which says that all Fourier modes decay exponentially. It is known that the Ott-Antonsen (OA) ansatz is a particular solution of the Watanabe-Strogatz (WS) equation due to the invariance of the Poisson kernel \cite{strogatz}.

For studying synchronization phenomena in Josephson junction arrays, Watanabe and Strogatz considered a set of coupled phase oscillators and demonstrated partial integrability of the system \cite{ws}. 
The WS theory yields dimensionality reduction without invoking any special ansatz. 
It consists of a coordinate transform that reduces the $N$ coupled system to a set of three equations accompanied by $N-3$ constants of motion. 
The WS theory is also valid for a finite-sized ensemble of oscillators, unlike the OA method, which is valid only in the thermodynamic limit. 
It was also shown that the WS theory is inherently related to the M\"{o}bius transformation and the equation of motion follows the Riccati equation \cite{ws_comments}. 
The OA ansatz is linked to the M\"{o}bius transformation via the Poisson Kernel whose invariance justifies the validity of the ansatz. 
The parameters of the WS theory and those of the Kuramoto mean-field model have been shown to be related to each other \cite{Pikovsky_Rosenblum}. 
 WS theory has been studied for star networks to obtain the time evolution of the order parameter measuring the strength of the synchronization \cite{star}.
The generalization of WS theory to higher dimensions has also been done to study the vector models of Kuramoto oscillators \cite{vectorws}.
The group theoretic structure of the higher-dimensional M\"{o}bius group, combined with hyperbolic geometry on a unit ball, describes the dynamics of the Kuramoto model on a sphere \cite{strogatz1}. 
The formalism of WS theory has been extended for higher-order harmonics as well, in which
 the M\"{o}bius transformation has been used to investigate the basins of attraction demonstrating that its pole aligns with the basin boundaries of the Kuramoto oscillators \cite{PhysRevE.100.062210}.
  Watanabe-Strogatz approach has an edge over the popular Ott-Antonsen method, as it does not involve any special ansatz and also reduces mathematical complexity in deriving dynamical evolution of associated parameters. 
  
 In this work, using the framework of WS theory, we study the extension of the Kuramoto model incorporating 1-simplex (pairwise) and 2-simplex (triadic) interactions along with their higher harmonics. The difference between the Kuramoto models investigated earlier using WS theory and the models considered here in this article is that all the previous works, to our knowledge, have considered either only pairwise or only triadic interactions. In contrast, here, we consider both the pairwise and triadic interactions.
 For WS theory to be applicable, these models must follow a Riccati-type equation. 
We show that the dynamics of both models can be obtained by following the WS procedure. For uniformly distributed constants of motion, the dynamics of WS parameters should be the same as that of the mean-field parameter of the Kuramoto oscillators \cite{Pikovsky_Rosenblum}. We first apply this result to coupled Kuramoto oscillators with pairwise and triadic interactions and explore the associated basin boundaries by defining the corresponding M\"{o}bius transformation. In addition, we introduce a coupled Kuramoto model that includes higher harmonics and higher-order interactions and show how including higher harmonics changes the basin boundaries. All results agree with the standard OA approach in the limit of identical oscillators. 

The paper is organized as follows. Section ~(\ref{sec2}) introduces two models investigated in this paper and provides a brief review of the WS theory and M\"{o}bius transformation for the first harmonic, along with its extension to the higher harmonics. Section ~(\ref{sec3}) applies the extended WS theory for the reduction in the dimensionality of both models. Section ~(\ref{sec4}) presents results based on numerical simulations of the evolution of these low-dimensional WS equations.

 \section{Model and Methods}\label{sec2}
 In this section, we describe M\"{o}bius transformation and its extended higher harmonic generalization and show their relations to WS theory. 
 We define two models such that both follow the Riccati equation, which is essential for WS theory to apply. 
 M\"{o}bius transformation, which follows the Riccati equation, is a central part of the WS theory since the evolution of the parameters in the WS theory is derived from the M\"{o}bius transformation. 
The extension of the Kuramoto model for $N$ oscillators is given by 
  \begin{multline}\label{model}
    \dot{\theta_i} = \omega + \frac{K_1}{N} \sum_{j=1}^{N} \sin{(h\theta_j - h\theta_i)} \\ + \frac{K_2}{N^2} \sum_{j,k =1}^{N} \sin{(m\theta_j + n \theta_k  - h\theta_i)}, 
 \end{multline}
where, $h,m,n$ are integers such that $h = m+n$ which ensure the phase shift invariance of the Kuramoto model, and $i = 1,2,...N$. The intrinsic natural frequency of oscillators is denoted by $\omega$, while $K_1$ and $K_2$ are the coupling strengths corresponding to 1-simplex (pairwise) and 2-simplex (triadic) interactions, respectively.  
Introducing the mean-field as
\begin{equation}\label{mean-field}
    z_p = \frac{1}{N} \sum_{j=1}^{N} e^{i p \theta_j} = r_p e^{i \Phi_p}.
\end{equation}
Eq.~(\ref{model}) can be written 
\begin{equation}\label{Riccati}
    \dot \theta_j = \omega + Im [H(t) e^{-i h \theta_j}],
\end{equation}
where $H(t)$ is a function of the complex order parameter $z_p$ (Eq.~(\ref{mean-field})). The parameter $h$ is chosen such that Eq.~(\ref{model}) yields Riccati-type structure.
Here, we consider two models. The first model consists of pairwise ($\sin{(\theta_j-\theta_i)}$) and triadic ($\sin{(2 \theta_j - \theta_k - \theta_i)}$) interaction terms, the model which has been investigated in detail using the Ott-Antesnon approach. In the second model, we consider another form of the triadic interaction ($\sin{(\theta_j + \theta_k - 2\theta_i)}$). With this later form of the triadic interaction, if one incorporates regular pairwise interactions ($\sin{(\theta_j-\theta_i)}$), it will not follow the Riccati equation. Hence, we consider higher harmonics in the pairwise term ($\sin{(2\theta_j-2\theta_i)}$). Note that for the first model, $m=2, n=-1$. Hence, the mean-field form of the Kuramoto oscillators corresponding to the first model can be written in the form of Eq.~(\ref{Riccati}) as
\begin{equation}\label{model1}
    \dot{\theta_i} = \omega + Im[(K_1 z_1 +K_2 z_2 \bar{z}_1)e^{-i \theta_i}].
\end{equation}
Here, $z_1$ and $z_2$ represent the centroid of all the $N$ oscillator with phases $e^{i\theta}$ and $e^{i2\theta}$, respectively.
For the second model, $m=1,n=1$ in Eq.~(\ref{model}), which yields the mean-field form of the coupled Kuramoto oscillators as 
\begin{equation}\label{model2}
    \dot \theta_i = \omega + Im [(K_1 z_2 + K_2 z_1^2) e^{-2\theta_i}].
\end{equation}
The pairwise coupling function is a periodic function of $2\pi$ of the phase difference ($\theta_j-\theta_i$) that can incorporate several higher-order harmonics.

Next, to study the dynamics of the models defined by Eqs.~(\ref{model1}) and (\ref{model2}), it is pertinent to provide a brief overview of WS theory and its relation to the M\"{o}bius transformation.

 \paragraph{{The Watanabe-Strogatz theory:}} To explain the dynamics of $N$ ($N \geq 3$) overdamped Josephson junction arrays, Watanabe and Strogatz \cite{ws} used a set of transformations known as the WS transformation that reduces the $N$ dimensional dynamics to lower dimensions that has $N-3$ constants of motion \cite{ws,ws_comments, strogatz,pr1,Pikovsky_Rosenblum,bolun}. 
 Consider a system of ordinary differential equation for $N$ globally coupled identical oscillators described by
 \begin{equation}\label{general}
     \dot \theta_j = A + B \sin{\theta_j} + C \cos{\theta_j} \hspace{1cm} j = 1,2,...N,
 \end{equation}
where, $A,B$ and $C$ are functions of phases ${\theta_i} (i\neq j)$, and $2\pi$-periodic in each argument. 
Further, WS transformation is given by 
\begin{equation}\label{WST}
    \tan\left[\frac{\theta_j (t) - \Theta(t)}{2}\right] = \frac{1-\rho(t)}{1+\rho(t)} \tan \left[\frac{\phi_j - \Psi(t)}{2}\right], 
\end{equation}
where, $\Psi(t), \rho(t)$ and $\Theta(t)$ are WS parameters having their own evolution equations \cite{ws,Pikovsky_Rosenblum}, thereby reducing the dimensionality of $N$ differential equations to three.
The transformed phases $\phi_j$ where $j=1,....N$ form a set of constant of motion provided that $\theta_j$ follow Eq.~(\ref{general}), which can be rewritten into a more symmetrical form as
\begin{equation*}\label{flow}
    \dot \theta_j = f e^{i \theta_j} + \bar f e^{-i \theta_j} + g,
\end{equation*}
where $f$ and $g$ are complex and real functions, respectively. 
The coordinate transformation ~(\ref{WST}) can be expressed as the M\"{o}bius transformation which plays an essential role in the WS theory.

\paragraph{{Relation to the M\"{o}bius transformation:}}
The M\"{o}bius transformation consists of linear fractional transforms ($\mathbb{C} \rightarrow \mathbb{C}$) and it maps an open unit disk in a complex plane onto itself in a one-to-one fashion. 
The general form of the time-dependent M\"{o}bius transformation and its inverse can be written as 
\begin{equation}\label{M1}
    \mathcal{M}: \phi_j \rightarrow \theta_j(t), \hspace{0.5cm} e^{i\theta_j (t)} = \frac{ e^{i \phi_j+ i \psi (t)}+\alpha(t)}{1+ \bar{\alpha}(t) e^{i \phi_j+ i \psi (t)}},
\end{equation} 
\begin{equation}\label{M2}
    \mathcal{M}^{-1}: \theta_j(t) \rightarrow \phi_j, \hspace{0.5cm} e^{i\phi_j} = e^{-i \psi(t)} \frac{e^{i \theta_j(t)}-\alpha(t)}{ 1- \bar{\alpha}(t) e^{i \theta_j(t)}}.
    \nonumber
\end{equation} 
Here, {$\theta_j$} is the phase of the $j^{th}$ oscillator, while the complex parameter $\alpha(t)$ satisfies $|\alpha(t)| \leq 1$, and the parameter $\psi(t)$ is a rotation angle and is given as: 
\begin{equation*}
    \alpha(t) = \rho(t) e^{i \Theta(t)}, \hspace{1cm} \psi(t) = \Theta(t) - \Psi(t).
\end{equation*}
The phases $\theta_j(t)$ evolve according to a time-dependent M\"{o}bius group action on complex unit circle as
\begin{equation*}\label{mt}
    e^{i \theta_j(t)} = \mathcal{M}_t (e^{i\phi_j}).
\end{equation*}
 Following Eq.~(\ref{M1}), the WS parameters $\alpha(t)$ and $\psi(t)$ evolve as \cite{strogatz, PhysRevE.100.062210};

\begin{equation}\label{rf1}
    \dot \alpha = -\frac{\bar{H}(t)}{2} \alpha^2 +  i \omega \alpha + \frac{H(t)}{2},
\end{equation}
\begin{equation}\label{rf2}
    \dot \psi = \omega + \frac{i}{2} [\bar{H}(t) \alpha - H(t) \bar \alpha].
\end{equation}
 Eqs.~(\ref{rf1}) and (\ref{rf2}) describes the dynamics of WS parameters for single harmonic.
 The general equation of motion for $N$ identical Kuramoto phase oscillators obeys the Riccati equation as follows,
\begin{equation*}\label{geom}
    \frac{d}{dt} (e^{i\theta_j}) = i \omega e^{i \theta_j} + \frac{1}{2} [H(t) - H^{*}(t) e^{2i\theta_j} ].
\end{equation*}

\paragraph{Generalization to the higher harmonics:} The extension of the WS theory for $l-$ harmonics ($l \geq 2$) has been done by Gong and Pikovsky \cite{PhysRevE.100.062210} where the $N$ phase oscillators obey,
\begin{equation*}\label{HHM1}
    \frac{d}{dt} (e^{i l \theta_j}) = i l \omega e^{i l \theta_j} + \frac{l}{2} [H(t) - \bar{H}(t) e^{2i l \theta_j}].
\end{equation*} 
The corresponding M\"{o}bius transformation can be written as
\begin{equation}\label{mm1}
    e^{il \theta_j (t)} = \frac{ e^{i \phi_j+ i \eta (t)}+\zeta(t)}{1+ \bar{\zeta}(t) e^{i  \phi_j+ i \eta (t)}},
\end{equation}
where, $\zeta$ and $\eta(t)$ in Eq.~(\ref{mm1}) is complex and real WS parameter for higher harmonics, respectively, and $\phi_j$ is a set of constant of motion.
Following the same procedure mentioned in \cite{strogatz,PhysRevE.100.062210}, we obtain
\begin{equation}\label{lrf1}
    \dot \zeta = l \left[-\frac{\bar{H}(t)}{2} \zeta^2 + i \omega  \zeta + \frac{H(t)}{2}\right],
\end{equation}
\begin{equation}\label{lrf2}
    \dot \eta = l\left[\omega  + \frac{i}{2}(\bar{H}(t) \zeta - H(t) \bar \zeta) \right].
\end{equation}
Hence, Eqs.~(\ref{lrf1}) and (\ref{lrf2}) for pure $l$-harmonic coupling differs from Eqs.~(\ref{rf1}) and (\ref{rf2}) by a factor of $l$.

\section{Analytical Results}\label{sec3}
This section implements the WS formalism and investigates the equation of motion obtained by the M\"{o}bius transformation for two models. 
The constants of motion associated with the WS formalism are assumed to be uniformly distributed throughout.
\paragraph{{Model 1:}}
Eq.~(\ref{model1}) is the Kuramoto model with sinusoidal global couplings containing pairwise and triadic interactions for $N$ oscillators. 
Comparing Eq.~(\ref{model1}) with Eq.~(\ref{Riccati}) we get,
\begin{equation*}
    H(t) = K_1 z_1 + K_2 z_2 \bar{z}_1.
\end{equation*}
The magnitude of $z_1$ (i.e., $r_1$) denotes the degree of synchronization, while the argument of $z_1$ (i.e., $\Phi_1$) is the mean phase of the oscillators. The value of $r_1$ lies between 0 and 1, where $r_1=0$ denotes the uniform distribution of oscillators, and $r_1 =1$ corresponds to global synchronization, and the magnitude of $z_2$ measures the 2-cluster synchronization in the Kuramoto model. Due to the higher-order term, $z_1$ does not completely characterize the model and depends on $z_2$. 
As constants of motion are uniformly distributed, the WS reduction scheme implies $\alpha$ to be the same as $z_1$.
Hence, we can directly insert the value of $H(t)$ in Eq.~(\ref{rf1}),
\begin{equation}\label{model1z}
    \dot{z}_1 = i \omega z_1 + \frac{1}{2}(K_1z_1+K_2 z_2\bar{z}_1) - \frac{1}{2}(K_1 \bar{z}_1 + K_2 \bar{z}_2 z_1) z_1^2.
\end{equation}
Substituting $z_p = r_p e^{i \Phi_p}$ in the Eq.~(\ref{model1z}), and separating real and imaginary parts, we get
\begin{multline}\label{dynamics1_1}
    \dot r_1 = \frac{K_1}{2} r_1 (1- r_1^2) + \frac{K_2}{2} r_1 r_2 \cos{(\Phi_2-2\Phi_1)}\\ -\frac{K_2}{2}r_2 r_1^3 \cos{(\Phi_2- 2\Phi_1)},
\end{multline}
\begin{equation}\label{dynamics1_2}
    \dot{\Phi}_1 = \omega + \frac{K_2}{2} r_2 \sin{(\Phi_2-2\Phi_1)} + \frac{K_2}{2} r_2 r_1^2 \sin{(\Phi_2-2\Phi_1)}.
\end{equation} 
It can be seen that the dynamics of $r_1$ is dependent on $r_2$ due to the triadic interactions. 
Eqs.~(\ref{dynamics1_1}) and ~(\ref{dynamics1_2}) match with the known results obtained through the OA approach in the limit of identical oscillators.
It can be noted that WS method reduces the complexity of the calculations only when the constants of motion associated with the M\"{o}bius transformation are uniformly distributed. As reflected by Eq.~(\ref{dynamics1_1}), $r_1$ will depend on $r_2$ for non-zero values of $K_2$.

\paragraph{Model 2:} The model for pairwise interactions of higher harmonics with higher-order interactions can be given by Eq.~(\ref{model2}) which
follows the Riccati equation, and hence the formalism is suitable for the model.
In the similar approach as followed for model~(\ref{model1}) in the previous part, comparing Eq.~(\ref{model2}) with Eq.~(\ref{Riccati}), we get
\begin{equation*}
    H(t) = K_1 z_2 + K_2 z_1^2.
\end{equation*}
The WS parameter $\zeta$ here is same as $z_2$ which evolves according to Eq.~(\ref{lrf1})
\begin{equation}\label{model2z}
    \dot{z}_2 = 2i \omega z_2 + (K_1z_2 + K_2 z_1 ^2) - (K_1\bar{z}_2 + K_2 \bar{z}_1^2)z_2^2.
\end{equation}
Again, putting $z_p = r_p e^{i \Phi_p}$ in Eq.~(\ref{model2z}), and separating real and imaginary parts, we get
\begin{multline}\label{dynamics2_1}
    \dot r_2 = K_1 r_2 (1-r_2^2) + K_2 r_1^2 (1-r_2^2) \cos{(2\Phi_1 - \Phi_2)},
\end{multline}
\begin{equation}\label{dynamics2_2}
     \dot \Phi_2 = 2\omega + K_2 r_1^2 \left(\frac{1+r_2^2}{r_2}\right) \sin{(2\Phi_1-\Phi_2)}.
\end{equation}
\begin{figure}[h]
    \centering
    \subfigure[First panel]{\includegraphics[width=0.8\linewidth]{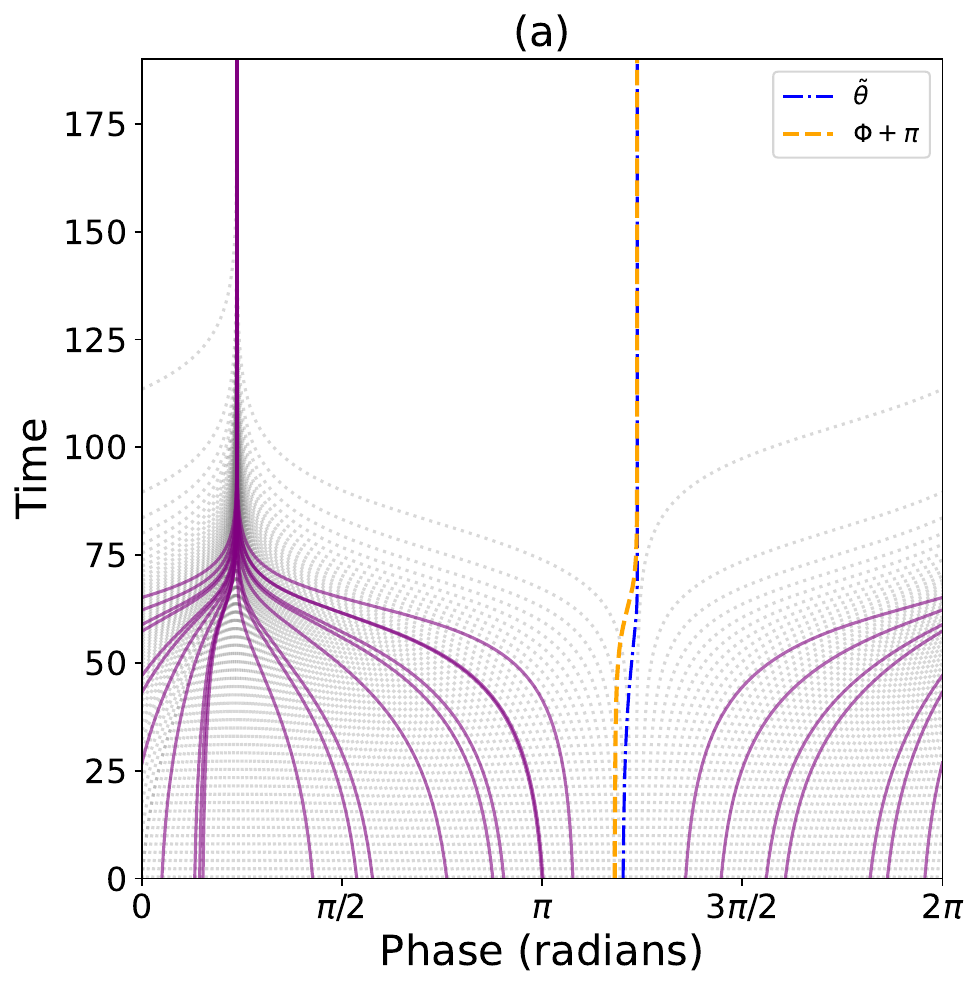}\label{Model1_1}}
    \subfigure[Second panel]{\includegraphics[width=0.8\linewidth]{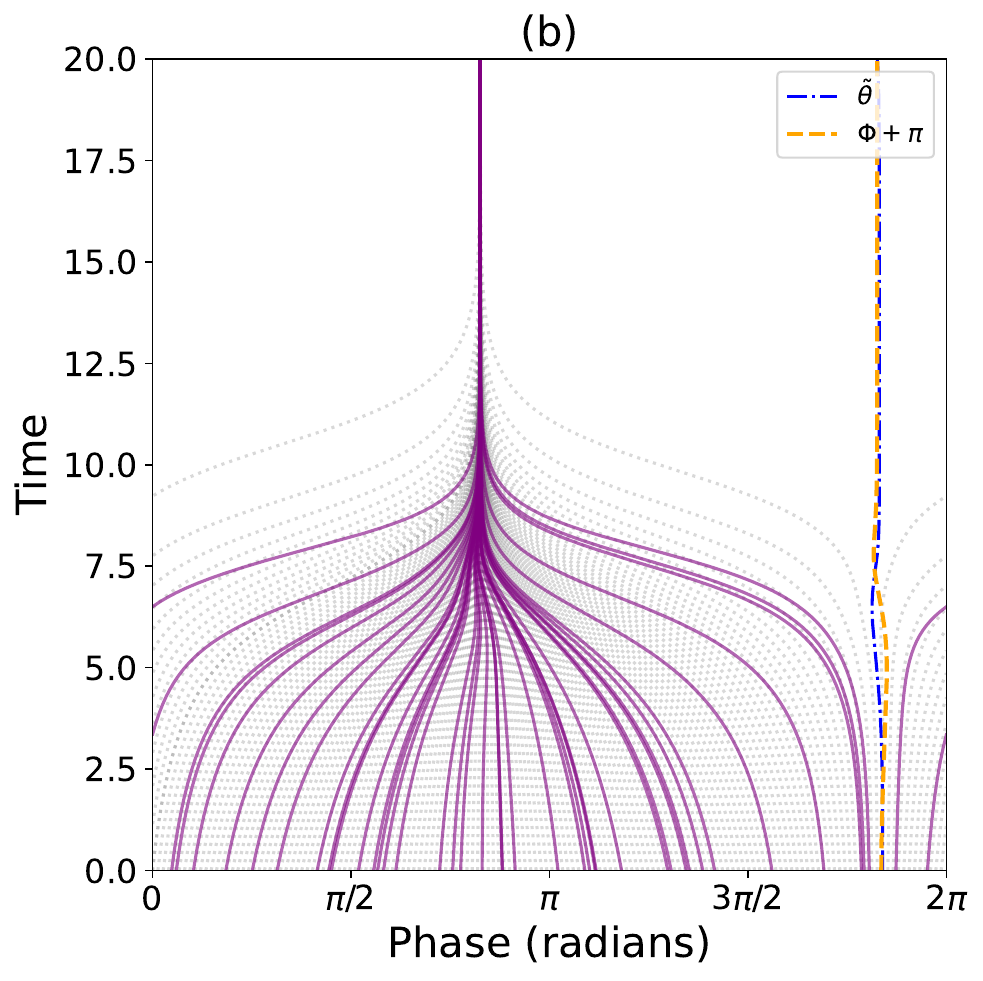}\label{Model1_2}}
    \caption{(Color online) Evolution of active phases, tracers and basin boundary as a function of time for model 1.
    Euler integration with step size $h=0.01$ of Eqs.~(\ref{WS_model1_1}) and (\ref{WS_model1_2})  with two different initial conditions. Gray dotted lines are tracers $\xi$, while purple solid lines indicate active phases. (a) $K_1 = 0.1$ and $K_2 = 0.1$ for $N=20$, (b) $K_1 = 0.5$ and $K_2 = 1.0$ for $N=40$. 
    Yellow dashed line refers the trajectory of WS parameter $\Phi + \pi$, while blue dot-dash line denotes the trajectory of tracer which ends up at singular point $\tilde{\theta} (\infty)$. Unstable trajectory is computed from Eq.~(\ref{hhm}) by using the final value of $\tilde{\phi}$ and instantaneous values of $\zeta(t)$ and $\psi(t)$ which were achieved while evolving Eqs.~(\ref{WS_model1_1}) and (\ref{WS_model1_2}) starting with a random set of initial conditions.}
    \label{Fig1}
\end{figure}

It can be seen from Eqs.~(\ref{dynamics2_1}) and (\ref{dynamics2_2}) that the dynamics of $z_2$ depends on $z_1$ due to the asymmetric part of the distribution. 
Proceeding further, we apply the self-consistency analysis 
\cite{Priyanka,Skardal} to characterize $z_1$. We note that, by changing the reference frame $\theta \rightarrow \theta + \omega_0 t$ and setting $\omega_0 = 0$, we get $\dot{\Phi}_1 = \dot{\Phi}_2 = 0$. 
Moreover, by rotating the initial condition $\theta(0) \rightarrow \theta(0) + \Phi_1(0)$ we set $\Phi_1 = \Phi_2 = 0$. 
For locked oscillators satisfying $|\omega| \leq K_1 r_2+K_2 r_1^2 $, there exist two stable fixed points $\theta^*(\omega)$ and $\theta^*(\omega) + \pi$ corresponding to two clusters. 
The value of $\theta^*(\omega)$ is given by
\begin{equation*}
    \theta^* (\omega)= \frac{1}{2} \arcsin{\left(\frac{\omega}{q}\right)}; \hspace{0.5cm} q = K_1 r_2 + K_2 r_1^2.
\end{equation*}
The density for the phase locked oscillator can be written as 
\begin{equation}\label{lock}
    f_{locked}(\theta,\omega,t) = \eta \delta(\theta-\theta^*) + (1-\eta) \delta(\theta-\theta^*-\pi),
\end{equation}
where, $\eta$ is the fraction of phase locked oscillators at $\theta = 0$, 
while drifting oscillators satisfying the condition $|\omega| > K_1 r_2+K_2 r_1^2 $ relax to stationary distribution
\begin{equation}\label{drift}
    f_{\text{drift}} = \frac{\sqrt{\omega^2-q^2}}{2\pi |\omega-q \sin{2 \theta}|}.
\end{equation}
In the continuum limit $z_1$ is given by the integral
\begin{equation*}
    z_1 = \int \int f(\theta,\omega,t) e^{i\theta} d\theta d\omega,
\end{equation*}
For locked oscillators Eq.~(\ref{lock}), we get 
\begin{equation}\label{locked_r1}
    r_1^{\text{locked}} = (2\eta -1 )\int_{-q}^{+q} e^{i \theta^*} g(\omega) d\omega,
\end{equation}
while, for drifting oscillators, we get from Eq.~(\ref{drift}) 
\begin{equation*}
    r_1^{\text{drift}} = \int_{|\frac{\omega}{q}|>1}\int_{-q}^{+q} e^{i \theta} f_{\text{drift}}(\theta,\omega) g(\omega) d\omega d\theta.
\end{equation*}
It is clear that $f_{\text{drift}} (\theta,\omega) = f_{\text{drift}}(\theta+\pi;\omega)$, and due to this symmetry, the contribution of the drifting oscillators goes to zero. Hence, value of $r_1$ solely depends upon the phase locked oscillators which turns out to be $r_1 = 2\eta - 1$ for identical oscillators. 
Next, Eqs.~(\ref{dynamics2_1}) and (\ref{dynamics2_2}) can be written as:
\begin{multline}\label{result2_1}
    \dot r_2 = K_1 r_2 (1-r_2^2) + K_2 (2\eta-1)^2 (1-r_2^2) \cos{(2\Phi_1 - \Phi_2)},
\end{multline}
\begin{equation}\label{result2_2}
     \dot \Phi_2 = 2\omega + K_2 (2\eta-1)^2 \left(\frac{1+r_2^2}{r_2}\right) \sin{(2\Phi_1-\Phi_2)}.
\end{equation}
Eq.~(\ref{locked_r1}) indicates that $z_1$ measures the asymmetry between two groups \cite{Skardal}, and hence $r_1 =0$ implies that the clusters are symmetrically distributed. Also, what follows from Eq.~(\ref{result2_1}) is that for $K_2=0$, the dynamics of $r_2$ is identical to Eq.~(\ref{dynamics1_1}) without the triadic interactions. We will get the same equation for $r_2$ as achieved for $r_1$ (Eq.~\ref{dynamics1_1}). For an identical system, $r_1$ becomes constant, and the dynamics will be characterized by  $r_2$ only (Eqs.~(\ref{result2_1}) and (\ref{result2_2})).

\begin{figure}[b]
    \centering
    \begin{subfigure}
        \centering
        \includegraphics[width=0.8\linewidth]{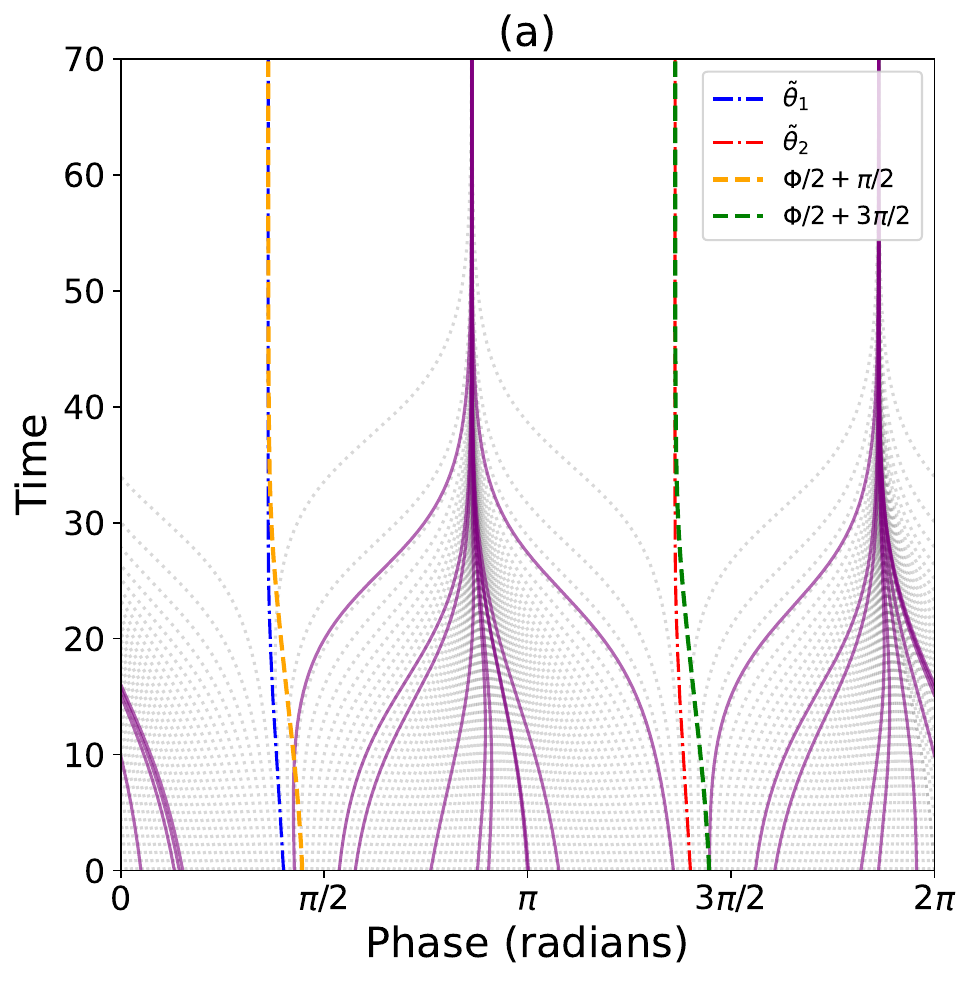}
        \label{Model2_1}
    \end{subfigure}
    \begin{subfigure}
        \centering
        \includegraphics[width=0.8\linewidth]{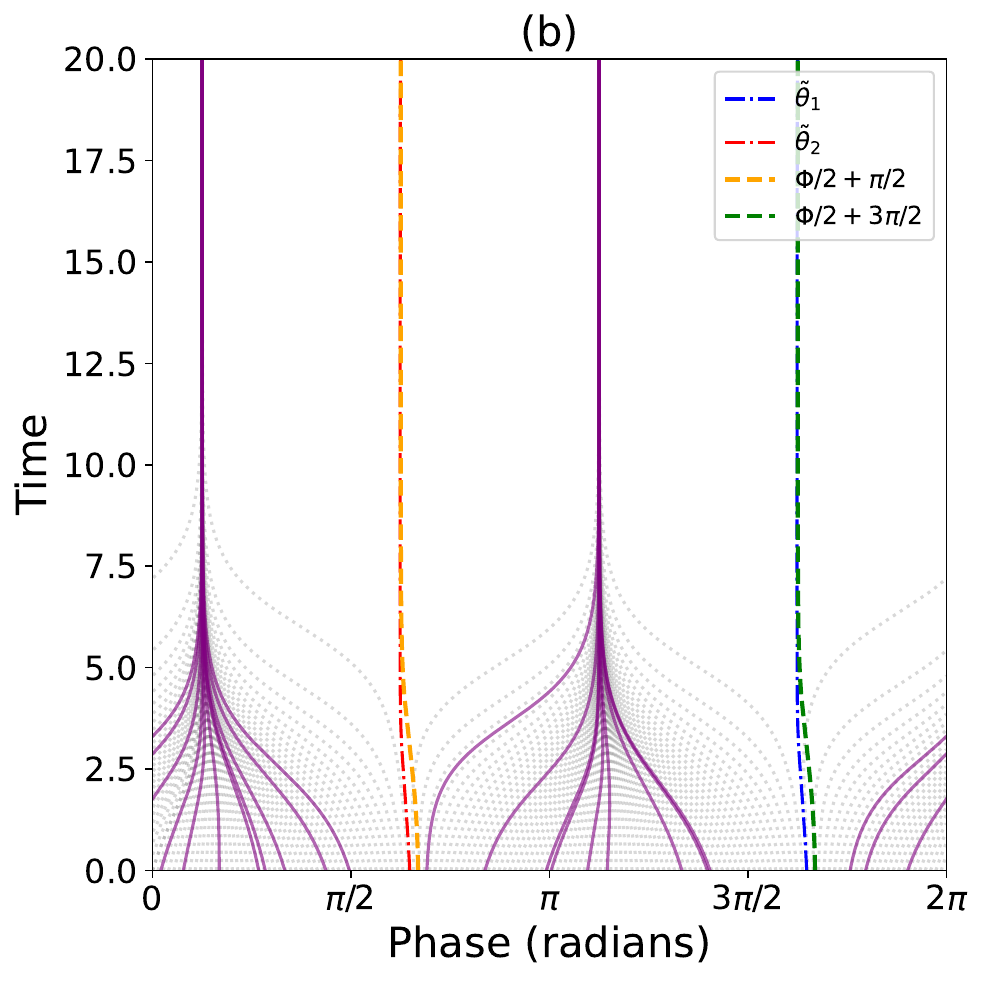}
        \label{Model2_2}
    \end{subfigure}
    \caption{(Color online) Evolution of active phases, tracers and basin boundaries as a function of time for model 2. Euler integration WS Eqs.~(\ref{WS_model2_1}) and (\ref{WS_model2_2}) with $h=0.01$ for two different initial conditions. Gray dotted lines are tracers $\xi$, while purple line indicates the active phases. (a) $K_1 = 0.1$ and $K_2 = 0.1$ for $N=20$, (b) $K_1 = 0.5$ and $K_2 = 0.5$ for $N=20$. 
    Yellow dashed and green dashed lines refers to the trajectories of WS variable $\Phi/2 + \phi/2$ and $\Phi/2 + 3 \pi / 2$, while blue and red dot-dash lines denotes the trajectories of two tracers which end up at singular points $\tilde{\theta}_1 (\infty)$ and $\tilde{\theta}_2 (\infty)$, respectively. The unstable trajectory is again computed as for Fig.~1 but by using Eqs.~\ref{WS_model2_1} and \ref{WS_model2_2}.
    }
    \label{Fig2}
\end{figure}
\section{Numerical Simulations}\label{sec4}
This section analyzes the evolution of basin boundaries by numerically solving Eqs.~(\ref{lrf1}) and ~(\ref{lrf2}).
 WS theory predicts that $l-$harmonic coupling yields the evolution of $l$ clusters ~\cite{PhysRevE.100.062210}. 
For uniformly distributed initial $\theta_i$,   $\phi$ can take different values. Therefore, for $|\zeta| < 1$, $\theta_i$ are all distinct, and $l-$cluster formation can occur only at the limit $|\zeta| \to 1$.
For the $l-$ harmonic, there exist $l$ basins of attraction corresponding to $l$ clusters. These basins of attraction form different boundaries separating different clusters, and change dynamically with time with the evolution of oscillators. It has been shown \cite{PhysRevE.100.062210} that these boundaries correspond to the pole of the M\"{o}bius transformation and therefore relate to the singularity that occurred in the WS transformation.
We rewrite Eqs~.(\ref{M1}) and (\ref{mm1}) in a single generalized form as
 \begin{equation}\label{hhm}
     e^{i m \tilde{\theta} (t)} = \frac{ \exp{[i \tilde{\phi}+ i \psi (t)]}+\zeta(t)}{1+ \zeta^{*}(t) \exp{[i \tilde{\phi}+ i \psi (t)]}},
 \end{equation}
 where, $m = 1, 2 $ corresponds to the model ~(\ref{model1}) and ~(\ref{model2}), respectively. 
Following the same procedure as mentioned in Ref.~\cite{PhysRevE.100.062210}, the transformation ~(\ref{hhm}) becomes singular at the limit $|\zeta| \rightarrow 1$. Taking $|\zeta|=1$, all values of $\phi$ are mapped in one-to-one fashion to the cluster states $\theta = \Phi / m + 2n\pi /m$, where $n=0,1,...m-1$, except when $\tilde{\phi} = \Phi + \pi - \psi$, where $\Phi = \text{arg} (\zeta)$. In the limit of $|\zeta| \rightarrow1^-$, this singularity is mapped by Eq.~(\ref{hhm}) onto the basin boundaries as evolution approaches $t \rightarrow \infty$: $\tilde{\theta} = \Phi / m + (2n+1)\pi /m$. Here, we take $\omega=0$ in Eqs.~(\ref{model1}) and ~(\ref{model2}) throughout the section.
\begin{figure}[t]
    \centering
    \begin{subfigure}
        \centering
        \includegraphics[width=0.8\linewidth]{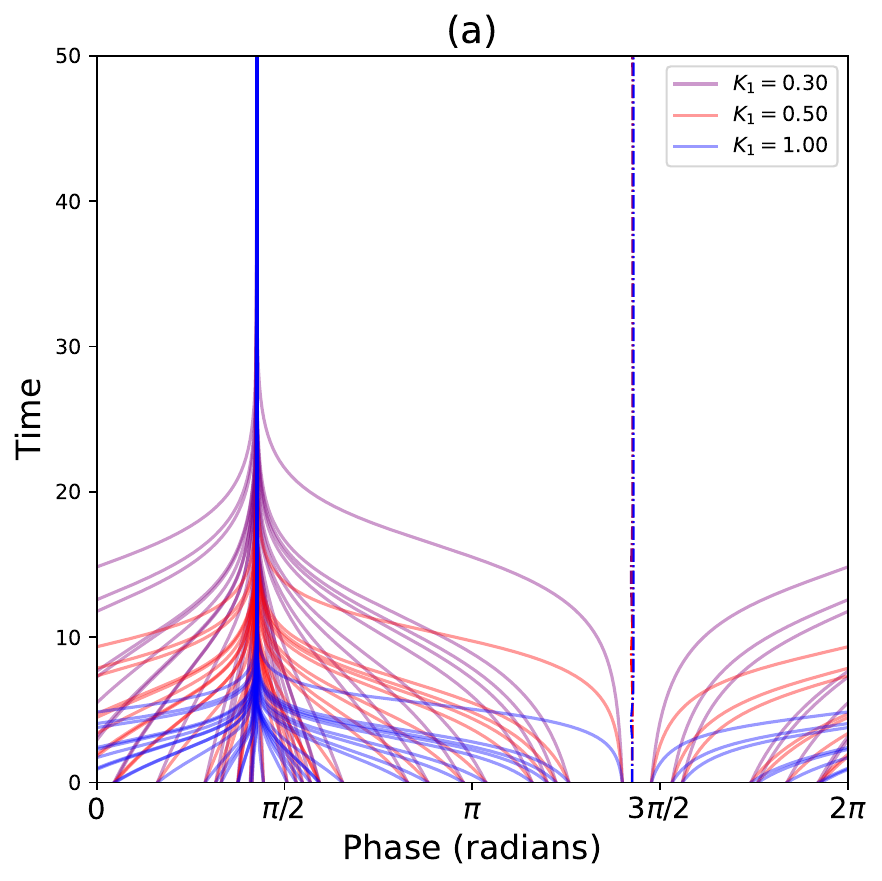}
        \label{Model1_1_K1_fix}
    \end{subfigure}
    \begin{subfigure}
        \centering
        \includegraphics[width=0.8\linewidth]{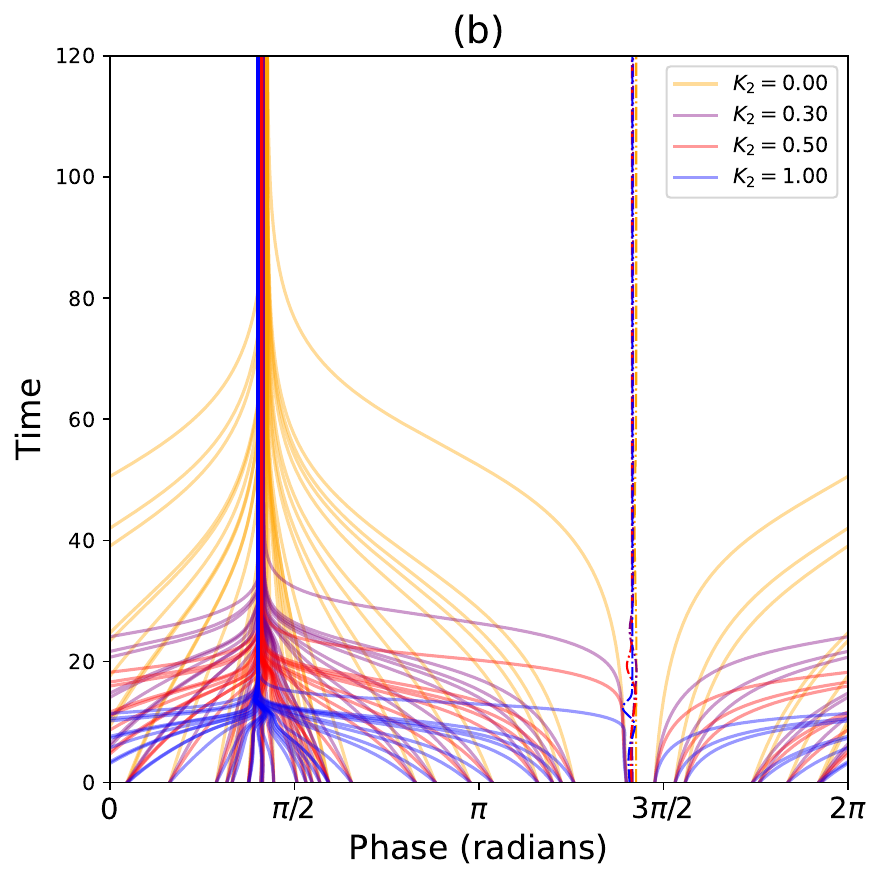}
        \label{Model1_1_K2_fix}
    \end{subfigure}
    \caption{(Color online) Effect of $K_1$ and $K_2$ on evolution of active phases (solid lines) and basin boundary (dot-dash line) for model 1 (Eqs.~(\ref{WS_model1_1} and \ref{WS_model1_2})) for $N=35$ oscillators. (a) $K_2=0.1$ is fixed, while $K_1$ has several different values, (b) $K_1$ is fixed to $0.1$ and $K_2$ takes different values for the same initial condition. }
    \label{Fig3}
\end{figure}
\textbf{$m=1$ case:} The system forms a single cluster within which oscillators settle without crossing the basin boundary (Fig.~(\ref{Fig1})). 
 $\tilde{\theta} = \Phi (\infty) + \pi$ maps to the basin boundary at $t \rightarrow \infty$ when $|\zeta(\infty)|=1$. 
The basin boundary can be identified at any given time by tracing these states backward over time. 

\begin{figure}[t]
    \centering
    \begin{subfigure}
        \centering
        \includegraphics[width=0.8\linewidth]{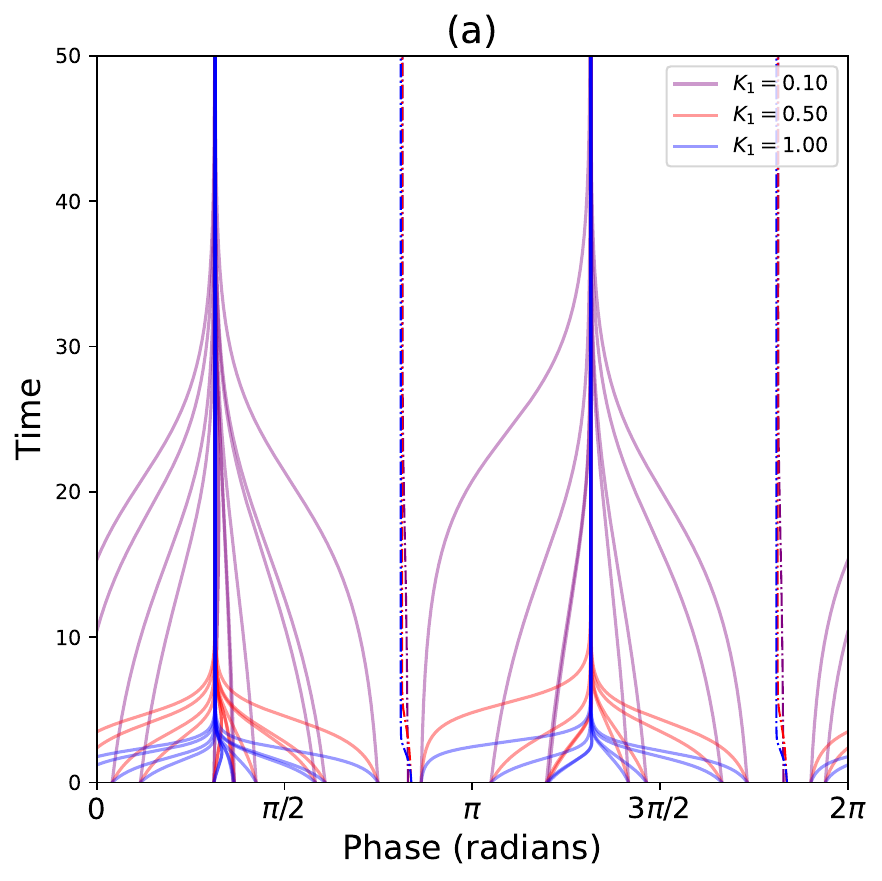}
        \label{Model2_1_K1_fix}
    \end{subfigure}
    \begin{subfigure}
        \centering
        \includegraphics[width=0.8\linewidth]{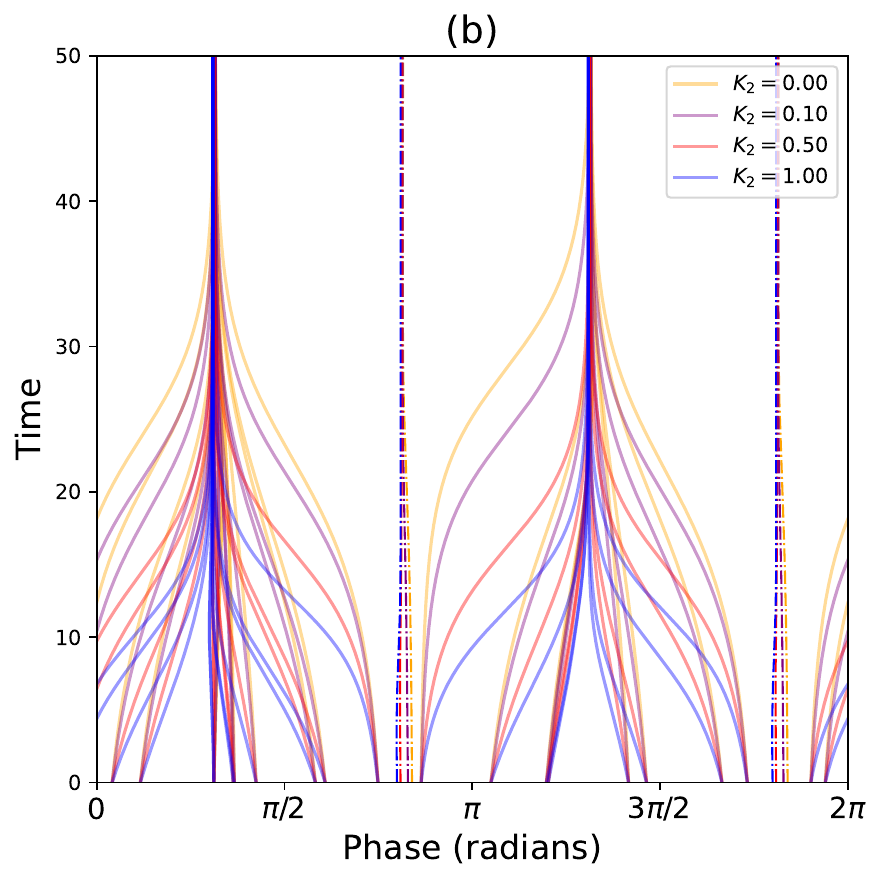}
        \label{Model2_1_K2_fix}
    \end{subfigure}
    \caption{(Color online) Effect of $K_1$ and $K_2$ on evolution of active phases (solid lines) and basin boundary (dot-dash line) for model 2 (Eqs.~(\ref{WS_model2_1} and \ref{WS_model2_2})) for $N=20$ oscillators. (a) $K_2=0.1$ is fixed, while $K_1$ has several different values, (b) $K_1$ is fixed to $0.1$ and $K_2$ takes different values for the same initial condition.}
    \label{Fig4}
\end{figure}
We can use another method to measure the value of $\tilde{\theta}$ as a function of time if we know the constant of motion $\tilde{\phi}$. However, $\tilde{\phi}$ can be evaluated at $t \rightarrow \infty$ ($\tilde{\phi} = \Phi(\infty) - \psi(\infty) + \pi$). 
Starting with $\theta_i$ randomly drawn from a uniform distribution on a circle, the model~(\ref{model1}) can be written in terms of the order parameter as
\begin{multline*}\label{model1_z}
     \dot{\theta}_i = K_1 |z_1| \sin{(\text{arg}(z_1)-\theta_i)} +\\ K_2 |z_2| |\bar{z}_1|\sin{(\text{arg}(z_2)- \text{arg}(z_1) - \theta_i)}.
\end{multline*}
We define passive ``tracers" $ \xi$ such that they are influenced by the global field which depends on the ``active" phases $\theta_i$, but do not themselves contribute to the global field, 
\begin{equation}\label{tracers}
   \dot{\xi} = \text{Im}[H(t) e^{-i \xi}],
\end{equation}
where, $H(t) = K_1 z_1 + K_2 z_2 \bar{z}_1$. 
The tracers can take any value between $0$ and $2\pi$. 
The motivation to introduce passive tracers as mentioned in Eq.~(\ref{tracers}) is to ``test" the strength of the field around the entire circle. 
For model ~(\ref{model1}), the WS parameters evolve as
\begin{equation}\label{WS_model1_1}
    \dot{\zeta} = \frac{1}{2}\left(K_1 z_1 +K_2 z_2 \bar{z}_1\right) - \frac{1}{2} (K_1 \bar{z}_1 + K_2 \bar{z}_2 z_1) \zeta^2, 
\end{equation}
\begin{equation}\label{WS_model1_2}
    \dot{\psi} = \text{Im} [(K_1z_1+K_2 z_2 \bar{z}_1) \bar{\zeta}].
\end{equation}
For numerical simulations, we evaluate Eqs.~(\ref{WS_model1_1}) and ~(\ref{WS_model1_2}) numerically. The initial values are chosen as $\zeta(0) = K_1z_1 (0)+K_2 z_2(0) \bar{z}_1(0)$ and $\psi(0) = 0$ for a given value of $K_1$ and $K_2$. 
Under this condition, the second WS equation (\ref{WS_model1_2}) becomes $\dot{\psi} = 0$ at time $t=0$. 
This choice of the initial condition is referred as natural condition as it depends on the initial phases \cite{PhysRevE.100.062210}. 
We solve Eqs.~(\ref{WS_model1_1}) and ~(\ref{WS_model1_2}) by using the Euler integration method. 
The time step for the numerical integration has been taken until the synchronized cluster is formed. 
Fig.~(\ref{Fig1}) denotes the evolution of phases randomly drawn for $N=20$ and $N=40$ for different values of $K_1$ and $K_2$ to demonstrate dependence on the choice of initial conditions. 
As $K_1$ and $K_2$ increase, the time required for the formation of a single cluster decreases for a set of given initial condition. 

\textbf{$m=2$ case:} The system forms two clusters for which there exist two distinct basin boundaries as shown in Fig.~(\ref{Fig2}). The unstable points $\tilde{\theta}_1 = \Phi (\infty) / 2 + \pi /2$ and $\tilde{\theta}_2 = \Phi (\infty)/ 2 + 3\pi /2$ are mapped at $t \rightarrow \infty$ as mentioned before. The evolution of these boundaries can be determined by the constant of motion $\tilde{\phi}$ at $t \rightarrow \infty$ ($\tilde{\phi} = \Phi(\infty) - \psi(\infty) + \pi$). The evolution of $\tilde{\theta}$ can be explored by using Eq.~(\ref{hhm}).

Further, writing model~(\ref{model2}) in terms of order parameter
\begin{multline*}\label{model2_z}
    \dot{\theta}_i = K_1 |z_2| \sin{(\text{arg}(z_2)-2\theta_i)} + \\K_2 |z_1|^2 \sin{(2 \text{arg}(z_1)-2\theta_i)}.
\end{multline*}
The tracer $\xi$ follows Eq.~(\ref{tracers}) with $H(t) = K_1z_2 + K_2 z_1^2$. 
For this case, the WS parameter evolves as
\begin{equation}\label{WS_model2_1}
    \dot{\zeta} = (K_1z_2 + K_2 z_1^2) - (K_1 \bar{z}_2 + K_2 \bar{z}_1^2) \zeta^2,
\end{equation}
and 
\begin{equation}\label{WS_model2_2}
    \dot{\psi} = 2 \text{Im}[(K_1z_2 + K_2 z_1^2)\bar{\zeta}].
\end{equation}
To perform numerical integration of Eqs.~(\ref{WS_model2_1}) and (\ref{WS_model2_2}), we consider the same initial conditions as taken for the previous case ($\zeta(0) = K_1z_2(0) + K_2 z_1^2(0) $ and $\psi(0) = 0$) for a given set of $K_1$ and $K_2$.
Fig.~(\ref{Fig2}) shows the evolution of the trajectories for $N=20$ for different values of $K_1 $ and $K_2$ implying two different sets of initial conditions. Again, the time taken for the formation of two groups decreases as $K_1$ and $K_2$ increase for the same set of initial conditions.

Next, Figs.~(\ref{Fig3}) and ~(\ref{Fig4}) manifest the effect of $K_1$ for a fixed $K_2$, and vice-versa for models ~(\ref{model1}) and ~(\ref{model2}), respectively. As $\theta_i$ starts with the same set of initial conditions, with an increase in $K_1$ and $K_2$, the time required to achieve synchronization, which we will refer to as transient time, decreases.
It should also be noted that since the pole of the M\"{o}bius transformation emerges only at $t \rightarrow \infty$, the initial basin boundaries cannot be determined without tracking the dynamics to the final state, as pointed out in~\cite{PhysRevE.100.062210}.

\section{Conclusion and Discussions}
Using WS formulation, we have analyzed the dynamical evolution of two different Kuramoto oscillator models by unifying them under a single framework. The only constraint here is that the model should obey the Riccati-type equation. We studied the Kuramoto model, which contains higher-order interactions and harmonics. The results match the standard Ott-Antonson approach in the limit of identical oscillators ($\omega = \omega_0 - i \Delta$, where $\Delta \rightarrow 0$).
We have explicitly studied phase evolution over time for these two models containing higher-order interactions following generalized M\"{o}bius transformation.
Although the form of the M\"{o}bius transformation remains unchanged in these models, the form of the coupling terms involving $K_1$ and $K_2$ significantly influences the evolution of the M\"{o}bius parameters. The numerical integration shows an excellent agreement with the WS equations and transformed phases.
 It should be noted that both the WS and the OA formulations are not suitable for mixed harmonics coupling.

One of the straightforward future directions of this study could be analyzing the behavior of asymmetrical clusters \cite{PhysRevE.100.062210}. Also, it is interesting to study the dynamical evolution via coupled M\"{o}bius maps \cite{mobiusmap}. Here, we have only focused on identical oscillators. Pikovsky and Rosenblum have developed the extended WS theory for the non-identical frequency distribution \cite{WS_nonidentical}, which can be probed further to incorporate the models presented here. It will also be interesting to study the effects of M\"{o}bius group on the formation of chimera \cite{chimera}.


\section{Acknowledgement}
SJ gratefully acknowledges SERB Grant No.
SPF/2021/000136.

\bibliographystyle{plain}
\bibliography{references}

\begin{thebibliography}{10}

\bibitem{Review}
Juan~A. Acebr\'on, L.~L. Bonilla, Conrad~J. P\'erez~Vicente, F\'elix Ritort, and Renato Spigler.
\newblock The kuramoto model: A simple paradigm for synchronization phenomena.
\newblock {\em Rev. Mod. Phys.}, 77:137--185, Apr 2005.

\bibitem{bolun}
Bolun Chen.
\newblock Dimensional reduction for identical kuramoto oscillators: A geometric perspective, ph.d. thesis.
\newblock {\em Boston College}, 2017.

\bibitem{star}
Hong-Bin Chen, Yu-Ting Sun, Jian Gao, Can Xu, and Zhi-Gang Zheng.
\newblock Order parameter analysis of synchronization transitions on star networks.
\newblock {\em Frontiers of Physics, 12, 120504}, 2017.

\bibitem{BEC}
Juan~J. Garc\'{\i}a-Ripoll, V\'{\i}ctor~M. P\'erez-Garc\'{\i}a, and Pedro Torres.
\newblock Extended parametric resonances in nonlinear schr\"odinger systems.
\newblock {\em Phys. Rev. Lett.}, 83:1715--1718, Aug 1999.

\bibitem{ws_comments}
Charles~J. Goebel.
\newblock Comment on “constants of motion for superconductor arrays”.
\newblock {\em Physica D: Nonlinear Phenomena,}, 80(Issues 1–2):18--20, 1995.

\bibitem{PhysRevE.100.062210}
Chen~Chris Gong and Arkady Pikovsky.
\newblock Low-dimensional dynamics for higher-order harmonic, globally coupled phase-oscillator ensembles.
\newblock {\em Phys. Rev. E}, 100:062210, Dec 2019.

\bibitem{mobiusmap}
Chen~Chris Gong, Ralf Toenjes, and Arkady Pikovsky.
\newblock Coupled m\"obius maps as a tool to model kuramoto phase synchronization.
\newblock {\em Phys. Rev. E}, 102:022206, Aug 2020.

\bibitem{riccati}
Cesar~A. {Gómez S.} and Alvaro Salas.
\newblock Special symmetries to standard riccati equations and applications.
\newblock {\em Applied Mathematics and Computation}, 216(10):3089--3096, 2010.

\bibitem{chimera}
Vladimir Ja\'{c}imovi\'{c} and Aladin Crnki\'{c}.
\newblock Möbius group actions in the solvable chimera model.
\newblock {\em International Journal of Modern Physics B}, 39(04):2540005, 2025.

\bibitem{JI}
Peng Ji, Jiachen Ye, Yu~Mu, Wei Lin, Yang Tian, Chittaranjan Hens, Matjaž Perc, Yang Tang, Jie Sun, and Jürgen Kurths.
\newblock Signal propagation in complex networks.
\newblock {\em Physics Reports}, 1017:1--96, 2023.

\bibitem{cosmology}
James~E Lidsey.
\newblock Cosmic dynamics of bose–einstein condensates.
\newblock {\em Classical and Quantum Gravity}, 21, 2004.

\bibitem{strogatz1}
Max Lipton, Renato Mirollo, and Steven~H. Strogatz.
\newblock The kuramoto model on a sphere: Explaining its low-dimensional dynamics with group theory and hyperbolic geometry.
\newblock {\em Chaos: An Interdisciplinary Journal of Nonlinear Science}, 31(9):093113, 09 2021.

\bibitem{vectorws}
M~A Lohe.
\newblock Higher-dimensional generalizations of the watanabe–strogatz transform for vector models of synchronization.
\newblock {\em Journal of Physics A: Mathematical and Theoretical}, 51(22):225101, may 2018.

\bibitem{strogatz}
Seth~A. Marvel, Renato~E. Mirollo, and Steven~H. Strogatz.
\newblock Identical phase oscillators with global sinusoidal coupling evolve by möbius group action.
\newblock {\em Chaos: An Interdisciplinary Journal of Nonlinear Science}, 19(4):043104, 10 2009.

\bibitem{Priyanka}
Bhuwan Moyal, Priyanka Rajwani, Subhasanket Dutta, and Sarika Jalan.
\newblock Rotating clusters in phase-lagged kuramoto oscillators with higher-order interactions.
\newblock {\em Phys. Rev. E}, 109:034211, Mar 2024.

\bibitem{ndiaye}
M.~Ndiaye.
\newblock The riccati equation, differential transform, rational solutions and applications.
\newblock {\em Applied Mathematics, 13, 774-792}, 2022.

\bibitem{OA}
Edward Ott and Thomas~M. Antonsen.
\newblock Low dimensional behavior of large systems of globally coupled oscillators.
\newblock {\em Chaos: An Interdisciplinary Journal of Nonlinear Science}, 18(3):037113, 09 2008.

\bibitem{Pikovsky_Rosenblum}
Arkady Pikovsky and Michael Rosenblum.
\newblock Partially integrable dynamics of hierarchical populations of coupled oscillators.
\newblock {\em Phys. Rev. Lett.}, 101:264103, Dec 2008.

\bibitem{WS_nonidentical}
Arkady Pikovsky and Michael Rosenblum.
\newblock Dynamics of heterogeneous oscillator ensembles in terms of collective variables.
\newblock {\em Physica D: Nonlinear Phenomena}, 240(9):872--881, 2011.

\bibitem{pr1}
Arkady Pikovsky and Michael Rosenblum.
\newblock Dynamics of globally coupled oscillators: Progress and perspectives.
\newblock {\em Chaos: An Interdisciplinary Journal of Nonlinear Science}, 25(9):097616, 07 2015.

\bibitem{Priyanka2}
Priyanka Rajwani and Sarika Jalan.
\newblock Stochastic kuramoto oscillators with inertia and higher-order interactions.
\newblock {\em Phys. Rev. E}, 111:L012202, Jan 2025.

\bibitem{rosu}
Haret~C. Rosu.
\newblock One‐parameter darboux transformations in thermodynamics.
\newblock {\em AIP Conference Proceedings}, 643(1):471--475, 11 2002.

\bibitem{Schuch_2014}
Dieter Schuch.
\newblock Nonlinear riccati equations as a unifying link between linear quantum mechanics and other fields of physics.
\newblock {\em Journal of Physics: Conference Series}, 504(1):012005, apr 2014.

\bibitem{Skardal}
Per~Sebastian Skardal and Alex Arenas.
\newblock Abrupt desynchronization and extensive multistability in globally coupled oscillator simplexes.
\newblock {\em Phys. Rev. Lett.}, 122:248301, Jun 2019.

\bibitem{kuramoto}
Steven~H. Strogatz.
\newblock From kuramoto to crawford: exploring the onset of synchronization in populations of coupled oscillators.
\newblock {\em Physica D: Nonlinear Phenomena}, 143:1--20, 2000.

\bibitem{tanaka}
Hisa-Aki Tanaka, Allan~J. Lichtenberg, and Shin'ichi Oishi.
\newblock First order phase transition resulting from finite inertia in coupled oscillator systems.
\newblock {\em Phys. Rev. Lett.}, 78:2104--2107, Mar 1997.

\bibitem{ws}
Shinya Watanabe and Steven~H. Strogatz.
\newblock Constants of motion for superconducting josephson arrays.
\newblock {\em Physica D: Nonlinear Phenomena}, 74(3):197--253, 1994.

\end{thebibliography}

\end{document}